\documentclass[11pt]{article}

\usepackage{mathptmx}
\usepackage[T1]{fontenc}

\usepackage{graphicx}
\usepackage[a4paper,margin=1in]{geometry}	% Including figure files
\usepackage{authblk} % per gestione elegante autori/affiliazioni
\usepackage{amsmath}	% Advanced maths commands
\usepackage{amssymb}	% Extra maths symbols

\usepackage{tabularx, array}
\usepackage{adjustbox}
\usepackage[outdir=./]{epstopdf}

\usepackage{float}
\usepackage{txfonts}
\usepackage[authoryear]{natbib}
\usepackage{longtable}

\usepackage{rotating}
\usepackage{lscape}
\usepackage{txfonts}
\usepackage{xcolor}
\usepackage{booktabs}
\usepackage[hidelinks]{hyperref}
\usepackage[capitalise]{cleveref}

\usepackage{authblk}

\def\kms{$\rm km\;s^{-1}$}

\def\Al*{Al$^*$}

\title{SPAN: A cross-platform Python GUI software for optical and near-infrared spectral analysis}

\author[1]{D. Gasparri\thanks{Corresponding author: \texttt{daniele.gasparri@postgrados.uda.cl}}}
\author[1]{L. Morelli}
\author[2,3]{U. Battino}
\author[4,5]{J. M\'endez-Abreu}
\author[4,5]{A. de Lorenzo-C\'aceres}

\affil[1]{\small Instituto de Astronomía y Ciencias Planetarias, Universidad de Atacama, Copayapu 485, 1530000 Copiapó, Atacama, Chile}
\affil[2]{\small Department of Physics, University of Naples Federico II, Via Cintia, Naples 80126, NA, Italy}
\affil[3]{\small INAF – Osservatorio Astronomico d’Abruzzo, Via M. Maggini, 64100 Teramo, Italy}
\affil[4]{\small Instituto de Astrof\'\i sica de Canarias, C/ V\'ia L\'actea s/n, E-38200 La Laguna, Spain}
\affil[5]{\small Departamento de Astrof\'\i sica, Universidad de La Laguna, C/ Astrof\'isico Francisco S\'anchez, E-38205 La Laguna, Spain}

\date{19 December 2025}

\begin{document}

\maketitle

 \begin{abstract}
The increasing availability of high-quality optical and near-infrared spectroscopic data, as well as advances in modelling techniques, have greatly expanded the scientific potential of spectroscopic studies. However, the software tools needed to exploit this potential often remain fragmented across multiple specialised packages, requiring scripting skills and manual integration to handle complex workflows.
In this paper we present \textsc{SPAN} (SPectral ANalysis), a cross-platform, Python-based Graphical User Interface (GUI) software that integrates the essential steps of the modern spectroscopic workflow within a single, user-friendly environment. SPAN provides a coherent framework that unifies data preparation, spectral processing, and analysis tasks, using the pPXF software as its core engine for full spectral fitting.
SPAN allows users to extract 1D spectra from FITS images and datacubes, perform spectral processing (e.g. Doppler correction, continuum modelling, denoising), and carry out detailed analyses, including equivalent width measurements, stellar and gas kinematics, and stellar population studies. It runs natively on Windows, Linux, macOS, and Android, and is fully task-driven, requiring no prior coding experience.
We validate SPAN by comparing its output with existing pipelines and literature studies. By offering a flexible, accessible, and well integrated environment, SPAN simplifies and accelerates the spectral analysis workflow, while maintaining scientific accuracy.
\end{abstract}

\section{Introduction}
The manipulation and analysis of astronomical spectra play a fundamental role in astrophysics, providing profound insights into the physical properties and evolutionary processes of stars and galaxies. 

Over the years, several software packages and pipelines have been developed to facilitate spectral manipulation and analysis. Some of these are general-purpose tools designed mainly for spectroscopic data reduction and manipulation. The widely used IRAF package \citep{tody1986,tody1993} has served as a standard reference for astronomers for decades. More recently, modern graphical user interface (GUI)-based software such as SPLAT-VO \citep{saloun2016} and pyspeckit \citep{ginsburg2022} have been developed to provide more updated alternatives. Other specialised tools focus on modelling galaxy spectra, such as PYMORPH \citep{vikram2010}, or galaxy spectral energy distributions (SEDs), such as Galapy \citep{ronconi2024}. Other packages are dedicated to automated spectral extraction, like ASPIRED \citep{lam2023}, to stellar spectra modelling and atmospheric parameter determination, like iSPEC \citep{blanco2014} or spectral visualisation and redshift determination such SpecPro \citep{masters2011}.

For both stars and galaxies, equivalent width (EW) measurements of absorption lines in the optical and near-infrared have long been essential in characterising their physical properties and stellar populations  \citep[e.g.][]{faber1973,burstein1984,worthey1994,trager2000}. More recent studies have expanded and refined these analyses in the optical \citep[e.g.][]{schiavon2007,morelli2008,morelli2012,mcdermid2015,molina2018} and near-infrared \citep[NIR, e.g.][]{molina2004,alton2018,riffel2019,morelli2020,baratella2020,gasparri2021,gasparri2024}. 
Different codes focus on line-strength EW measurements, including the PACCE code \citep{riffel2011}, the Fortran-based LECTOR code \citep{vazdekis2011}, the C++-based Indexf code \citep{cardiel2010}, and ROBOSPEC \citep{waters2013}.

Full spectral fitting algorithms provide a complementary and robust approach to the analysis of stellar and galaxy spectra. Rather than focusing on specific spectral features, these methods simultaneously fit the spectrum with a combination of template spectra, allowing for a more comprehensive extraction of physical information. For integrated spectra of galaxies, full spectral fitting has proven particularly effective for deriving the kinematics of both stellar and gaseous components, as well as for reconstructing the star formation history (SFH) \citep[e.g.][]{cappellari2004, wilkinson2015, mcdermid2015, pessa2023}.

Several algorithms have been developed for the full spectral fitting technique to study the stellar populations of galaxy spectra. The pPXF code \citep[][and references therein]{cappellari2023} is widely used for extracting stellar kinematics, stellar population properties, and SFH. Additional full spectral fitting algorithms and pipelines include STARLIGHT \citep{fernandes2005}, STECMAP \citep{ocvirk2006}, VESPA \citep{tojeiro2007}, ULySS \citep{koleva2009}, FADO \citep{gomes2017}, FIT3D \citep{sanchez2016,lacerda2022}, and FIREFLY \citep{wilkinson2017}, some of which incorporate machine-learning techniques \citep[e.g.][]{fabbro2018}.

Several pipelines have been developed to perform advanced spectral analysis of galaxies. Pipe3D is specialised for the analysis of CALIFA, MaNGA, and SAMI data \citep{sanchez2016a} based on the FIT3D fitting tool. The GIST pipeline \citep{bittner2019,bittner2021} integrates pPXF and GANDALF \citep{sarzi2017} for the spectral analysis of stars and gas from galaxy datacubes. The updated nGIST pipeline \citep{fraser2024} focuses on MUSE datacube analysis.

Despite the strengths of these software packages in their respective domains, a gap remains in the availability of a fully cross-platform general purpose software that integrates all the modern spectral manipulation and analysis tools within an intuitive GUI.

SPAN is a long-term project that began in 2020. It has been developed to provide users with a simple and intuitive GUI that integrates the most commonly used tools for the manipulation and analysis of astronomical spectra, with a specific focus on unresolved spectra of galaxies. The primary goal of SPAN is to simplify and make the spectral manipulation and analysis more efficient by offering a single, fully cross-platform, and user-friendly environment. At the same time, SPAN is designed to be flexible and extensible, allowing users to modify and enhance the source code to introduce new routines or adapt existing ones to the evolving needs of the astronomical community.

In this paper, we present version 6.6.X of SPAN\footnote{\texttt{SPAN} and the documentation are available at \url{https://pypi.org/project/span-gui/}}, and provide a comprehensive overview of its capabilities and operation. 
The structure of the paper is as follows. In \cref{sec:description}, we introduce the main functionalities of SPAN. In \cref{sec:architecture} we describe the structure and architecture of the source code. In \cref{sec:spman}, we focus on the spectral manipulation tasks supported by the software, while in \cref{sec:span} we present the available spectral analysis tools. In \cref{sec:performances}, we compare the results produced by SPAN with other widely used software and evaluate its performance. In \cref{sec:examples} we present a science case, focusing on the spiral galaxy NGC~1097 using publicly available MUSE data. Planned future developments are outlined in \cref{sec:future}, and our conclusions are summarised in \cref{sec:conclusion}.

\section{General description}
\label{sec:description}
SPAN is a graphical software designed for the manipulation and analysis of large samples of one-dimensional (1D) spectra, enabling the efficient processing of hundreds or thousands of spectra within a single interactive framework. The input 1D spectra, can be obtained independently by the users or extracted by SPAN from fully reduced 2D long-slit and 3D (i.e. datacube) FITS files. The 1D spectra are loaded into the main interface via a spectra list file, which is a simple ASCII text file containing the paths and filenames of the input spectra.

\begin{figure}
\centering
\includegraphics[width=14truecm]{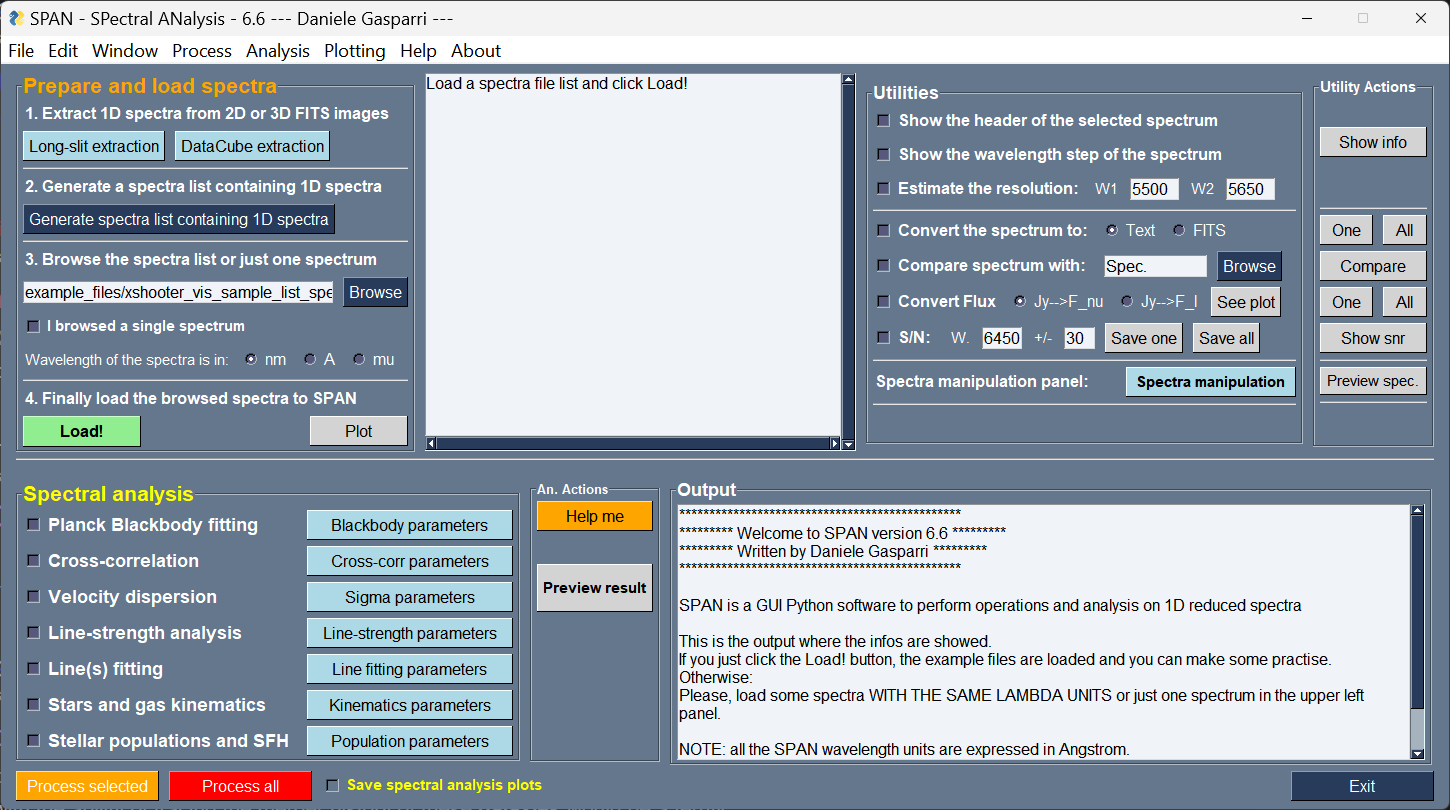}
\caption{The main panel of SPAN.}
\label{fig:span_gui}
\end{figure}

The main graphical panel of SPAN (\cref{fig:span_gui}) provides access to all available operations, or tasks, through an intuitive and user-friendly interface. Two processing modes are supported: in the \texttt{Process selected} mode, the currently selected spectrum from the list is processed using the active tasks; in the \texttt{Process all} mode, the same set of user-defined parameters and tasks is automatically applied to all loaded spectra.

SPAN includes a wide range of tasks for both spectral manipulation and analysis. The `Utilities' frame allows users to retrieve information on the loaded spectra (e.g. FITS header, wavelength step, resolution) and perform operations such as continuum normalisation, rebinning, resolution degradation, smoothing, and Doppler and redshift correction, through the `Spectra manipulation' panel.
The analysis of the spectra is performed in the `Spectral Analysis'  frame, which provides access to several techniques commonly applied in stellar and extragalactic spectroscopy. These include black-body fitting, cross-correlation, velocity dispersion measurements, line fitting, kinematics, stellar population and line-strength analysis based on EW measurement of spectral indices. Among the available full spectral fitting algorithms, SPAN employs pPXF, which is widely regarded as one of the most efficient and reliable methods for recovering the kinematics and the stellar population properties of galaxy spectra \citep{woo2024}. 
Considering the purpose of SPAN, we chose not to include Bayesian inference tools, such as Bagpipes \citep{carnall2018} or Prospector \citep{johnson2021}, whose computationally intensive calculations and long runtimes are not suitable for the analysis of large spectral datasets.

\begin{figure}
\centering
\includegraphics[width=8truecm]{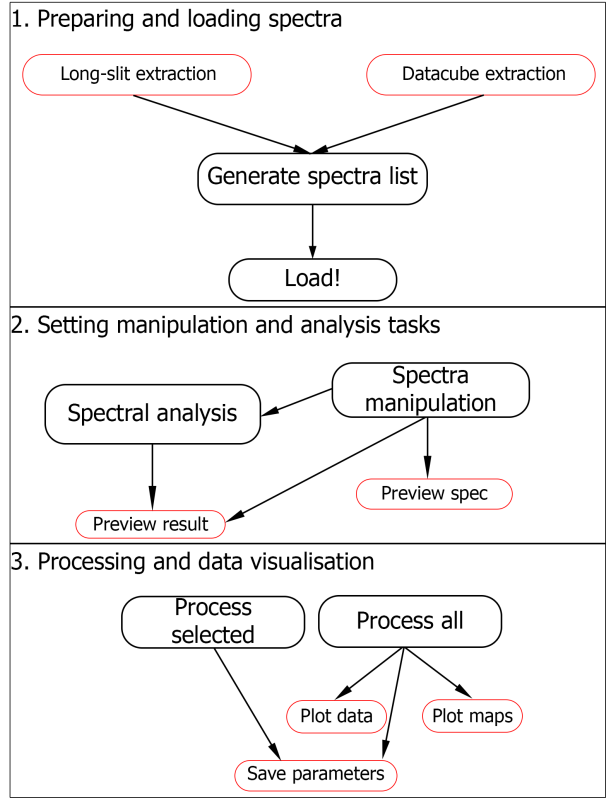}
\caption{Workflow diagram of SPAN from data extraction to spectral analysis and visualisation. In black we marked the required steps, while in red we marked the optional steps. Black rectangular frames delineate the three sequential blocks that compose a complete workflow.}
\label{fig:span_flux}
\end{figure}

\Cref{fig:span_flux} summarises the typical SPAN workflow, with required steps in black and optional ones in red. The workflow consists of three sequential blocks: (i) spectra preparation and loading, (ii) task configuration and parameter setup, and (iii) processing, visualisation, and parameter saving.

Users can start from reduced 2D FITS spectra or datacubes to extract the required 1D spectra. The `Long-slit extraction' sub-program allows users to extract 1D spectra from fully reduced and wavelength-calibrated long-slit FITS images, assuming the dispersion axis lies along the X direction.
Users can either extract a single spectrum by manually defining the spatial range to be integrated, or generate a series of spectra with a constant signal-to-noise ratio (S/N) by binning adjacent spatial rows. The extraction is performed through direct flux summation.
The `Cube extraction' sub-program is dedicated to extracting spectra from fully reduced 3D datacubes in FITS format. It can produce 1D spectra at a desired S/N using the Voronoi binning algorithm of \citet{cappellari2003}, or through manual selection of spatial regions to be combined. Optional features include masking, redshift correction, and custom definitions of the wavelength range used for S/N computation. SPAN natively supports datacubes from MUSE, CALIFA, WEAVE LIFU, and JWST NIRSpec IFU, automatically recognising their spatial and spectral axes. Other datacube formats can be handled by providing custom reading routines that can be easily loaded into the software.

Once the 1D spectra are loaded through the spectra list file, all the tasks are active. The preview modes are useful for adjusting the parameters. The process modes save the results. Only in `Process all' mode the results of the spectral analysis tasks are stored to disc. From these files, users can generate maps or plots with dedicated subprograms. Finally, the whole set of tasks and parameters can be saved before closing the program to be re-used and shared.

The output generated by SPAN is stored in a folder called \path{SPAN_results} whose location is set by the users. The output of the spectra manipulation tasks can be generated either in the `Process selected' and `Process all' mode. They are processed spectra saved in fits files containing the wavelength and flux values and stored in the \path{SPAN_results/spec} directory. The output generated by the spectral analysis frame consists of multiple ASCII files containing the results of the selected spectral analysis task and FITS auxiliary spectra (e.g., residuals and best fit templates of the full spectral fitting based tasks).

\section{Architecture}
\label{sec:architecture}

\subsection{The GUI}

The GUI has been built using the \path{FreeSimpleGui} module, a wrapper for the \path{Tkinter} \citep{lundh1999} framework. \path{FreeSimpleGui} is distributed under a GNU Lesser General Public License and is available on GitHub (\url{https://github.com/spyoungtech/FreeSimpleGUI}). The module is included in the SPAN distribution and does not require installation via \texttt{pip}.

The GUI has been designed to be as intuitive and clear as possible, featuring a minimalistic layout that ensures quick loading and refreshing. Tasks are activated via checkboxes next to their respective labels with a simple mouse click or a tap on touchscreen devices. Task parameters are accessible through a click or a tap to the relative buttons. 

SPAN automatically detects the operating system in use and dynamically adjusts the GUI structure to account for screen resolution and the scaling factor. The layout for Android has been specifically adapted to accommodate the 16:9 horizontal resolution and the smaller screen size of mobile devices. 

Since version 4, a menu bar with hotkey support has been introduced, providing quick access to subprograms, useful options (e.g., saving and loading user-defined parameters and task states), and additional documentation. To enhance usability on touchscreen devices running Android, the menu bar has been replaced with a set of buttons positioned at the top of the interface.

To improve user experience and maintain consistency across different operating systems, SPAN includes an embedded output window that simulates a terminal in Windows, Linux, and Android environments. For compatibility reasons, the macOS version retains a standard external terminal output.

In addition, tooltips, status messages and help buttons have been implemented to provide real-time help for any specific task selected, enhancing the accessibility and clarity of the interface. 

\subsection{Code structure}
The code is structured into multiple modules to ensure clarity and maintainability. Many of SPAN tasks rely on custom routines that have been employed in published works \citep{gasparri2021,gasparri2024}, and these are briefly described in \cref{sec:spman,sec:span}. To integrate external modules, such as pPXF, we have implemented dedicated wrappers that bridge the graphical input with the parameters required by these modules.

The Python routines specifically developed for SPAN, along with the wrappers that call the pPXF algorithm, are stored in the \path{span_functions} directory. These routines are organised into well-documented files grouped by their functionality. As an example, the \path{spec_manipul} module groups all routines dedicated to spectra manipulation.

The methods responsible for constructing GUI windows, reading spectra, handling events, storing parameters, and executing the functions for the specific tasks are located in the \path{span_modules} directory. These modules are further categorised, with each one of them managing a distinct component of the program. Input parameters are handled using Python's \path{dataclasses} library, introduced in Python 3.7. The dataclass structure ensures efficient storage, passing, and updating of GUI parameters across different SPAN modules.

Data handling and numerical operations leverage the \path{NumPy} library \citep{harris2020array} for computational efficiency. The flexibility of \path{Matplotlib} \citep{Hunter:2007} is exploited to produce clear, publication-quality plots while enabling direct user interaction with spectra (e.g., for mask creation). The \path{Astropy} \citep{astropy2013,astropy2018}, and \path{SciPy} \citep{scipy2020} libraries facilitate spectral manipulation and statistical analysis, respectively, while \path{Pandas} \citep{mckinney-proc-scipy-2010,reback2020pandas} is employed for managing input data and storing output information.

To enhance stability, robust exception handling has been implemented to prevent GUI crashes caused by invalid operations. When an error occurs, specific messages are displayed via pop-up windows and terminal output. 

The program status, including all activated tasks and any modified parameters, can be saved in .json files and restored at any time. This functionality allows users to create specific sets of tasks and parameters that can be applied to different datasets, facilitating easy reuse and sharing.

\section{The spectra manipulation panel}
\label{sec:spman}

\begin{figure}
\centering
\includegraphics[width=14truecm]{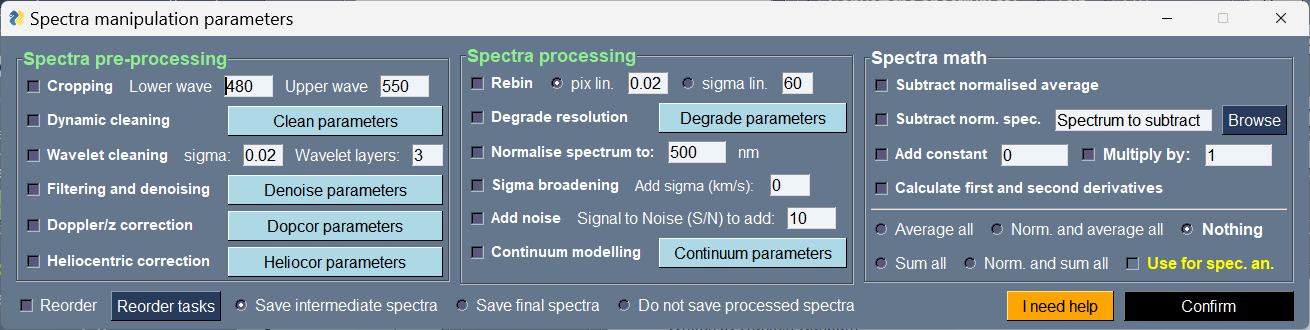}
\caption{The spectra manipulation panel of SPAN.}
\label{fig:spectra_manipulation}
\end{figure}

In many scientific cases, the 1D extracted spectra should be manipulated before being analysed. This manipulation could involve a wide range of operations, such as de-redshift, continuum modelling, filtering and denoising, and resolution degradation. The spectral manipulation panel of SPAN (\cref{fig:spectra_manipulation}) provides a set of tasks that allow users to perform such operations. The Python routines written for this panel are grouped in the \path{spec_manipul} module within the \path{span_functions} folder.

The spectra manipulation panel is divided into three sections:
Spectra Pre-Processing (e.g., cropping, cleaning, filtering, Doppler and heliocentric correction), Spectra Processing (e.g., rebinning, degrading resolution, velocity dispersion broadening, continuum modelling), and Spectra Math (simple operations on one or all spectra). 
By default, tasks within this panel are executed in serial mode. However, their order can be modified by the users. Tasks in this panel directly modify the input spectra, and the processed spectra can be used in further spectral analysis.

Among the most relevant tasks, the Doppler/z correction allows shifting spectra to the rest frame using the function \path{dopcor}. This can be done either by providing the redshift $z$ or, for local objects ($z < 0.01$), the radial velocity $v_r$ in \kms. A related task, `Heliocentric correction', corrects spectra for Earth's motion around the Sun using \path{helio_corr}, leveraging the \path{astropy.coordinates} module to compute the heliocentric velocity.

The Resolution degradation tool adjusts the spectral resolution, either in terms of resolving power $R$ or full width at half maximum (FWHM) using Gaussian convolution (\path{Gaussian1DKernel} from \path{astropy.convolution}). This function is critical when comparing spectra obtained with different instruments or theoretical models.

Another key task is `Continuum modelling', which provides two methods for the continuum parametrisation: a simple broad Gaussian convolution (\path{sub_cont}) to approximate the continuum for spectra with weak features, and a polynomial fitting approach (\path{continuum}), based on \path{Numpy.polyfit}, allowing also for masking emission or absorption lines.

Other available tasks include:
\begin{itemize}
    \item Cropping: Selects a wavelength range of the spectra (\path{crop_spec}).
    \item Denoising and filtering: Includes sigma clipping (\path{sigma_clip}), wavelet-based noise removal (\path{denoise_wavelet}), and various smoothing techniques (boxcar, Gaussian, Butterworth filters).
    \item Rebinning: Resamples spectra to a linear wavelength or logarithmic velocity scale (\path{resample}, \path{log_rebin}).
    \item Normalisation: Normalise flux relative to a user-defined reference wavelength (\path{norm_spec}).
    \item Sigma broadening: Applies Gaussian velocity dispersion broadening (\path{sigma_broad}).
    \item Add noise: Introduces Gaussian noise corresponding to a target S/N (\path{add_noise}).
    \item Basic mathematical operations: Includes stacking, averaging, subtraction, multiplication, and derivative calculations. 
\end{itemize}

\section{Spectral analysis}
\label{sec:span}
Spectral analysis is a powerful tool for extracting key information from the spectra of stars and galaxies. In extragalactic research, it serves as the primary method to uncover the properties and evolutionary history of galaxies. For this reason, in the last years many efforts have been made to develop algorithms to best retrieve this information, along with Simple Stellar Population (SSP) models based on empirical and theoretical spectral libraries. SPAN embeds the most widely used stellar population template libraries for galaxy studies: MILES \citep{vazdekis2010,falcon2011}, EMILES \citep{vazdekis2016}, sMILES \citep{knowles2023}, XSL \citep{verro2022}, GALAXEV \citep{bruzal2011} and FSPS \citep{conroy2010}. It also allows users to use their own set of template, if required.
In the following subsections, we present the spectral analysis tasks contained in SPAN and we focus on the most important that rely on the line-strength and full spectral fitting techniques.

\subsection{Basic analysis tools}
A series of tasks can be used to fit spectral lines (emission or absorption), the continuum shape of stellar spectra with a black-body function, to cross-correlate the spectra with a template to measure the radial velocity component or the redshift, and to fit prominent spectral features with a template or a series of Gaussian functions. These basic tasks use built-in function grouped in the module \path{spectra_analysis}. Here is a brief description of how they work:

\begin{itemize}
\item Planck blackbody fitting: This task fits the spectral continuum with a Planck blackbody function, providing an estimate of the colour temperature of the source. The fit is more reliable when the fitted wavelength range includes a sufficiently broad portion of the continuum, ideally encompassing the region near the peak or the curvature of the Planck function. The result is the best estimation of the effective temperature of the considered stellar spectrum and the best fit Planck function model;

\item Cross-Correlation: This task cross-correlates any user input template spectrum with the input spectra and measures the wavelength shift, both in terms of radial velocity and redshift. It uses the \path{cross_corr} function of the \path{spectra_analysis}. The result is the radial velocity or z of the spectrum. The uncertainty on the redshift (or velocity) is empirically estimated from the shape of the cross-correlation function. Specifically, the FWHM of the main correlation peak is measured, and the corresponding one sigma uncertainty is obtained as $\sigma = \mathrm{FWHM}/2.355$.

\item Velocity dispersion: This task performs a least-squares fit of a selected spectral region with a single template spectrum to estimate the velocity dispersion. It employs the \path{sigma_measurement} function of the \path{spectra_analysis} module.
Uncertainties are estimated through Monte Carlo simulations, by adding Gaussian noise consistent with the measured S/N on the pseudo continuum to the best-fit template. The procedure is repeated 100 times, and the standard deviation of the recovered velocity dispersion values is taken as the uncertainty.
Although less comprehensive than pPXF, which simultaneously fits multiple templates and higher-order kinematic moments, this task provides a robust and significantly faster estimate of the velocity dispersion. It is particularly useful for obtaining a first-order kinematic characterisation or when computational efficiency is a priority.

\item Line(s) fitting: This task allows to fit any absorption or emission line with a combination of a straight line for the continuum and a Gaussian function for the line. In addition, a dedicated routine automatically fits the three near-infrared Calcium Triplet (CaT) lines using the \path{cat_fitting} function, while general-purpose line fits are handled by the \path{line_fitting} function of the \path{spectra_analysis} module. The task returns the best-fitting model together with the integrated line flux EW.
\end{itemize}

\subsection{Lick/IDS line-strength analysis}
SPAN includes a dedicated option for automated line-strength analysis through the measurement of absorption-line EWs, following the Lick/IDS system of spectral indices originally defined in the 1980s on a sample of 460 stellar spectra \citep{burstein1984,faber1985,worthey1994,worthey1997}. These indices have become a cornerstone for the characterisation of unresolved stellar populations in galaxies, particularly for deriving luminosity-weighted estimates of age, metallicity, and $\alpha$-enhancement.

The Lick/IDS indices are defined over the 4000 -- 6000 \AA \ wavelength range and require a specific workflow before they can be measured and used for model comparisons. SPAN implements the full pre-processing pipeline necessary to bring the spectra into the Lick/IDS system and to estimate the stellar population parameters via model comparison. This includes:

\begin{itemize}
\item Emission-line subtraction, using the pPXF algorithm to simultaneously fit stellar and gaseous components.
\item Rest frame correction, based on the radial velocity derived during the fit.
\item Spectral convolution to match the resolution of the Lick/IDS system (FWHM = 8.4 \AA).
\item Velocity dispersion correction of the measured EWs, following the method and the coefficients of \citet{trager1998}.
\end{itemize}

SPAN measures the Lick/IDS indices using built-in passband definitions stored in the \path{system_files} directory, and proceeds to estimate the stellar population parameters through interpolation with theoretical SSP model grids, as widely performed in literature \citep[e.g.][]{trager2000,kuntschner2000,Mehlert2003, morelli2008,morelli2016}. SPAN includes the Lick/IDS line-strength measurements for the models of \citet{thomas2011}, MILES, sMILES, and XSL, stored in the \path{system_files} directory and ready to be called.

The interpolation with the model-based Lick/IDS indices is performed considering the H$\beta$ - [MgFe]$^{`}$ of \citet{thomas2003} and the <Fe> - Mgb index-index grids and uses the \path{lick_pop} function of the \path{spec_analysis} module. The interpolation between the measured and model indices is performed with two different approaches. The first one is a multi-dimensional linear interpolation that uses the \path{griddata} function of the \path{SciPy} module to retrieve the age, metallicity and $\alpha$-enhancement, where available, at the positions of the observed Lick/IDS indices on the index grids produced by the models. The second is an experimental approach that uses a supervised machine learning algorithm. The Gaussian Process Regression (GPR) in particular, has probed to be a powerful tool in retrieving parameters from stellar spectra \citep{bu2015,bu2020}. To our knowledge, no attempt has been made to apply the GPR to unresolved galaxy spectra and the Lick/IDS line-strength study. 

In SPAN, we trained the GPR algorithm with the line-strength predictions of the SSP models using the \path{scikit-learn} 1.4.2 module\footnote{If using a different version of the \path{scikit-learn} module, users should delete the embedded original trained models. Therefore, SPAN will perform an automatic training using the current version of the \path{scikit-learn} module the first time the task is called.}. The trained models, stored in the \path{system_files} folder, are used to constrain the age, metallicity, $\alpha$-enhancement and the relative uncertainties.  
The GPR method returns both the predicted values and their associated interpolation uncertainties, which reflect the confidence of the regression given the local density and coverage of the SSP model grid.

\begin{figure}
\centering
\includegraphics[width=14truecm]{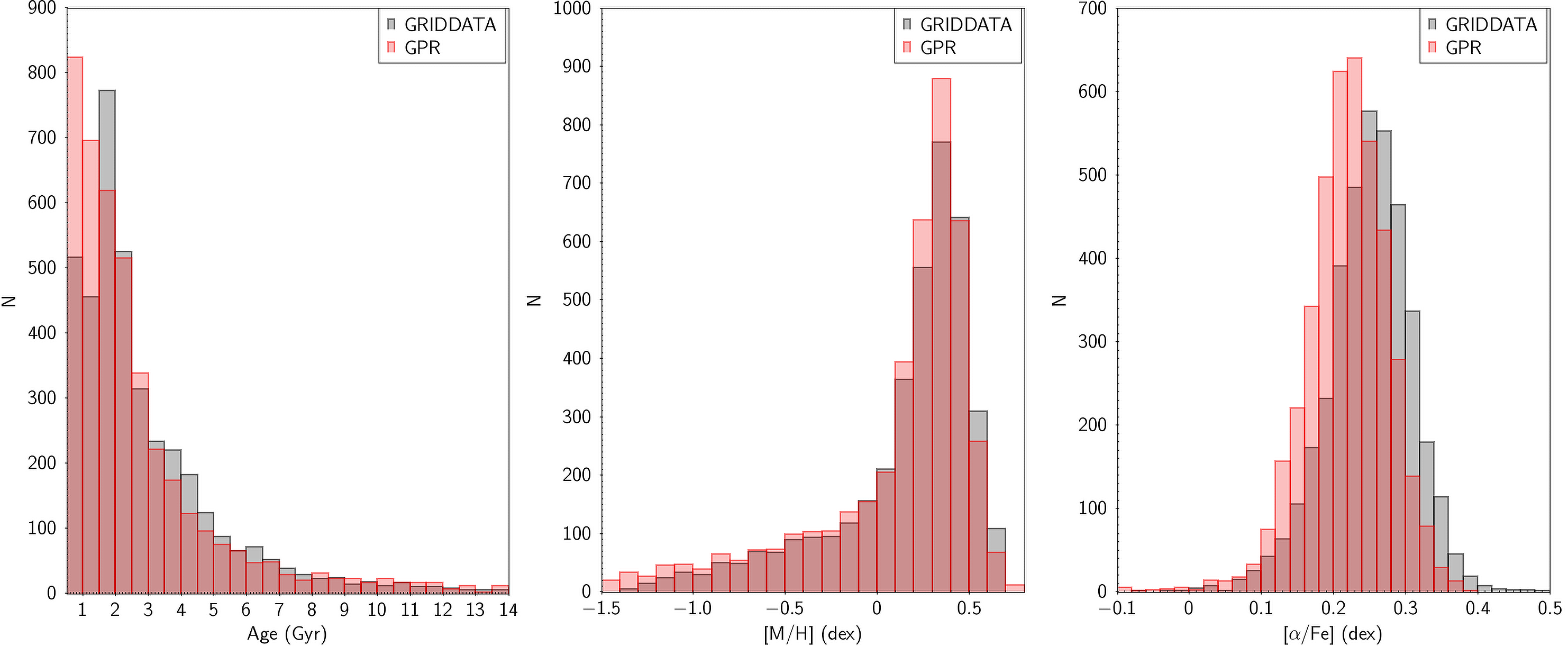}
\caption{Distribution of luminosity weighted age, metallicity and $\alpha$-enhancement measured with the Lick/IDS indices and the \citet{thomas2011} models for 4175 Voronoi bins of the galaxy NGC 1097  analysed in \cref{sec:examples}, between the multi-dimensional linear interpolation with the \texttt{griddata} function of the \texttt{SciPy} module (dark grey) and the machine learning GPR method (red).}
\label{fig:lick_SSP_comparison_interp}
\end{figure}

In \cref{fig:lick_SSP_comparison_interp} we compare the distribution of luminosity weighted age, metallicity and [$\alpha$/Fe] for 4175 Voronoi bins of the MUSE datacube of the galaxy NGC 1097 analysed in \cref{sec:examples}, using the multi-dimensional linear interpolation with the \texttt{griddata} function of the \texttt{SciPy} module (dark grey) and the machine learning GPR method (red), with the \citet{thomas2011} SSP models.  

We find good agreement between the two methods for metallicity. The age distributions are broadly consistent, although the GPR algorithm tends to slightly overestimate the youngest ages ($\sim$1~Gyr) and underestimate intermediate ages ($\sim$5--7~Gyr) relative to the linearly interpolated values. For [$\alpha$/Fe], the two distributions show a systematic offset of approximately 0.02~dex.

These preliminary results are promising, particularly because the GPR method offers two practical advantages over the simple linear interpolation. First, linear interpolation assumes locally linear behaviour and returns a single deterministic value, whereas the GPR model can follow smooth non-linear dependencies of the indices on stellar parameters and naturally provides a predictive uncertainty that quantifies the reliability of the interpolation itself. Second, the GPR is approximately an order of magnitude faster than the linear interpolation and the estimation of the relative uncertainties with Monte Carlo simulations and remains robust even for lower signal-to-noise spectra, where the linear interpolation often fails to converge. Further optimisation and testing of the GPR implementation are planned for future releases of SPAN.

\subsection{Custom line-strength analysis}

SPAN supports the EW measurement of user-defined line-strength indices. This option allows users to define and apply custom index definitions beyond standard systems such as Lick/IDS, thereby facilitating studies of specific absorption or emission features across different spectral ranges.

Custom indices are specified via a plain-text configuration file, where each column corresponds to an index definition. The first row contains the index names, followed by six rows indicating the wavelength intervals (in \AA): the blue pseudo-continuum (min, max), the red pseudo-continuum (min, max), and the line bandpass (min, max). These definitions can be loaded directly to SPAN. A single index option can be activated to measure or test an user-input index definition. 
The software returns the measured EWs both in \AA \ and magnitudes. Uncertainties are estimated via Monte Carlo simulations, following \citet{gasparri2021}.  For each spectrum and spectral index, SPAN generates 100 simulated pseudo-continua within the line bandpass limits, adding random noise consistent with the rms measured in the pseudo-continuum bands. The EW is then computed for each simulated continuum, which by definition should yield zero. The standard deviation of these EW values around zero provides a direct estimate of the uncertainty associated with the index measurement.

When dealing with galaxy spectra, it is often necessary to correct line-strength indices for the effects of velocity dispersion broadening. To support this, SPAN implements a two-step correction pipeline:
\begin{itemize}
\item Estimation of correction coefficients: A grid of broadened template spectra is created by convolving them to zero-velocity-dispersion spanning velocity dispersions from 50 to 400 \kms \ in 50 \kms \ steps. For each index, the EW is measured across the grid. A third-order polynomial is then fitted to the EW vs. $\sigma$ relation, providing the correction coefficients as a function of velocity dispersion. This method follows the prescription of \citet{trager1998}, and has been extended to arbitrary user-defined indices following the approach of \citet{cesetti2013} and \citet{morelli2020}.
\item Application of the correction coefficients: Once the correction coefficients are computed, the raw EWs measured on the spectra can be corrected as a function of their velocity dispersion. The task returns corrected EWs and provides the corresponding propagated uncertainties.
\end{itemize}

\subsection{Stars and gas kinematics}

The `Stars and gas kinematics' task is designed to extract the kinematic properties of both stellar and gaseous components using pPXF. The graphical wrapper provides access to a wide range of parameters and options that can be customised for various astrophysical contexts. The task supports the measurement of radial velocity, velocity dispersion, and higher order Gauss--Hermite moments (up to $h_6$). The spectra can be fitted using the built-in SSP template libraries or with any user-defined set of single FITS files (with wavelength in \AA) located in a local directory.

For the stellar kinematics, users may fit a single component or activate the two-component mode to model kinematically decoupled stellar populations. In this case, templates can be selected according to specific ages and metallicities, if available.

The gaseous component can be fitted in two ways: either simultaneously with the stellar component, or in a second step after fixing the stellar kinematics obtained from an initial fit. The latter approach is useful when stellar and gas components are separated in kinematic space. Emission-line fluxes are also measured and included in the output. 
 
pPXF allows the use of additive and multiplicative Legendre polynomials to correct for residual differences in the continuum shape between the observed spectrum and the templates (see \citealt{cappellari2023}). The additive polynomials modify the overall continuum level by adding low-order terms, while the multiplicative ones rescale the continuum shape without altering the relative flux of narrow spectral features. When accurate emission-line fluxes and EW are required, we recommend deactivating the additive polynomials (by setting their degree to --1) and using only multiplicative polynomials, as suggested by \citet{cappellari2023}, in order to preserve the flux calibration and ensure physically meaningful line intensities.

Dust attenuation can be included in the fit using the \texttt{dust} keyword (introduced in pPXF version 8.2.1, replacing the former \texttt{reddening} keyword). This option applies a wavelength-dependent attenuation curve to the model templates, introducing a free parameter that scales the amount of dust extinction during the fit. For the gaseous component, the \citet{calzetti2000} attenuation law is used by default, while the stellar continuum is attenuated using the \citet{cappellari2023} curve, which reproduces the mean extinction behaviour of nearby galaxies. In both cases, the \texttt{dust} keyword modifies the model flux as a smooth multiplicative function of wavelength, allowing the recovered line fluxes and continuum shape to account for internal extinction in a physically consistent manner.
Users can also apply custom masks to exclude specific wavelength regions from the fit, which is useful for removing spectral artefacts or poorly calibrated regions.

If the noise level of a spectrum is unknown or variable across the sample, an optional subroutine can automatically estimate the optimal noise level. This is achieved following the procedure described by \citet{mcdermid2015} and explained in pPXF documentation\footnote{\url{https://pypi.org/project/ppxf/}}. An initial unregularised fit (with the \texttt{bias} keyword set to zero) is performed, and the initial noise value is scaled to obtain $\chi^2/\mathrm{goodPixels.size} = 1$, where goodPixels.size is the number of pixels in the spectrum being fitted. 
By using the optimal noise estimation, the formal errors estimated by pPXF can be considered a fair representation of the real uncertainties, for sigma >  2 velscale, where sigma is the line-of-sight velocity dispersion and velscale is the velocity sampling of the input spectrum.
A more robust treatment of uncertainties on the derived kinematic parameters can be performed via Monte Carlo simulations, which can be configured and activated within the graphical interface. With this option, the bestfit template found is perturbed N times (where N is defined by the user) by adding Gaussian noise corresponding to the S/N given by pPXF. The perturbed templates are therefore fit with the same parameters but without regularisation (\texttt{bias} keyword of pPXF set to zero). The standard deviation of the kinematic moments measured on the simulated spectra is taken as representative of the uncertainty.

\subsection{Stellar populations and SFH}
\label{sec:kinematics}
The `Stellar populations and SFH' task is optimised to estimate the properties (age, metallicity, $[\alpha/Fe]$) of the stellar populations and the non parametric SFH on galaxy spectra. 

For this task we decided to include also a subset of the sMILES templates \citep{knowles2023}, compiled with Salpeter IMF and four values of $[\alpha/Fe]$, from $[\alpha/Fe]$ = -0.2 to $[\alpha/Fe]$ = +0.4. We have customised the wrappers to the pPXF package to let it work in a 3D space: age, metallicity, and $[\alpha/Fe]$, following the work of \citet{grasser2024}. The modifications have been placed in the module \path{build_templates} in the \path{span_functions} subfolder within the class \path{smiles}. Users can include any custom set of sMILES templates by simply replacing the content of the \path{spectralTemplates/sMILES_afeh} folder. 

Any user-defined template set consisting of single FITS files that follow the standard MILES file naming convention\footnote{\url{http://research.iac.es/proyecto/miles/pages/ssp-models/name-convention.php}} can be added and used directly within the GUI. Any set of templates can also be added by preparing the .npz file using the pPXF standard and loading it to SPAN.

By default, the templates are normalised to the V-band using the \path{norm_range} keyword of pPXF, and the weights given by the fitting procedure therefore represent the light fraction of each template. SPAN makes use of the output attribute \path{.flux} introduced in version 7.4.5 of pPXF to convert the light weights also to mass weights, without repeating the fit.

We also introduced two routines, one for the automated best noise estimation of each spectrum and another for determining both the best noise and regularisation parameter, following \citet{cappellari2017}. For the best noise estimation option, an unregularised fit is performed, then the initial noise guess is rescaled so that $\chi^2/\mathrm{goodPixels.size} = 1$.
If the option to determine the best regularisation parameter is activated, first it computes the best noise determination and then a series of regularised fits is iteratively executed, with the `Regul. error' = 1/regularisation refined using a bisection method until `Current delta chi2' converges within $\pm 10\%$ of `Desired delta chi2'. If convergence is not achieved within ten iterations, the initial `Regul. error' guess is retained. This approach is computationally intensive and is best suited for analyses involving fewer than 300 templates and limited wavelength range ($\Delta \lambda \lesssim 2000$ \AA).

The determination of uncertainties on the measured stellar parameters is performed by residuals bootstrapping, following \citet{kacharov2018} and using the wild bootstrap method of \citet{davidson2008} as suggested by \citet{cappellari2023}. 
The residuals of the regularised fit are multiplied by random factors of $\pm1$ with equal probability, then added back to the best-fitting model N times (where N is set by the users) to create N synthetic spectra. Each synthetic spectrum is fitted with the same configuration and a small amount of regularisation (regul = 5) as recommended by \citet{kacharov2018}. The dispersion of the recovered stellar population parameters over all bootstrap realisations provides an empirical estimate of their uncertainties.

Additionally, we included in this task also the option to determine stellar parameters using Lick/IDS index analysis as performed in the `Line-strength analysis' task. This feature leverages pPXF output by measuring Lick/IDS indices on emission-corrected spectra, considering Doppler/z shift and velocity dispersion values computed by pPXF.

\section{Tests and performances}
\label{sec:performances}
SPAN has been tested on Windows (10 and 11 64-bit versions), Linux (Debian, Fedora, Ubuntu), macOS, and Android (versions 11 to 14) machines, as well as with all Python 3.10+ versions.
The program has been used with IRAF-reduced spectra, SDSS spectra, the first release of the IRTF library by \citet{rayner2009}, and the extended version by \citet{villaume2017}, SAURON spectra, X-Shooter library spectra \citep{arentsen2019, gonneau2020, verro2022b}, JWST NIRSPEC extracted spectra, MUSE, CALIFA, WEAVE LIFU, and JWST NIRSpec datacubes, (E)MILES, GALAXEV, FSPS, and XSL stellar libraries. Additionally, it complies with the ESO standard for extracted spectra products.

\subsection{Results validation}
We thoroughly tested the main spectral analysis tasks of SPAN against different codes and reference datasets.   In particular, we compared the line-strength EW measurements obtained with SPAN against the tabulated E-MILES values in the LIS system at a constant resolution of FWHM\,=\,8.4\,\AA\ \citep{vazdekis2010}.  We used 424 E-MILES SSP models computed with BASTI isochrones and a Kroupa IMF, covering the metallicity range $-1.26 \leq \mathrm{[M/H]} \leq 0.26$.  To minimise discretisation effects, all spectra were resampled to a linear wavelength grid with $\Delta\lambda=0.1$\,\AA. 

\begin{figure}
\centering
\includegraphics[width=14truecm]{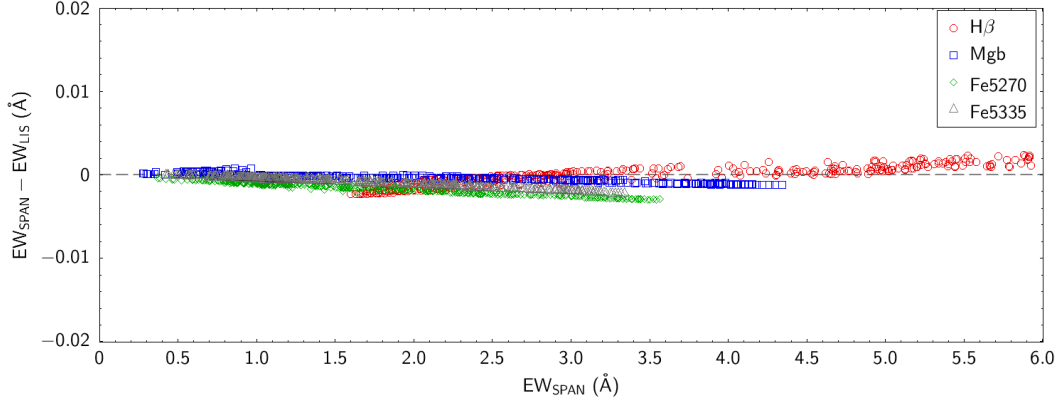}
\caption{Residuals between the EWs measured by SPAN and the tabulated E-MILES values for four optical line-strength indices widely used as age, metallicity, and [$\alpha$/Fe] diagnostics: H$\beta$, Mgb, Fe5270, and Fe5335.  
The dashed line marks the zero level, while the vertical extent of the Y axis corresponds to the typical uncertainty range expected at $\mathrm{S/N}\!\approx\!100$, as discussed by \citet{cardiel1998}.}
\label{fig:lick_emiles}
\end{figure}

\Cref{fig:lick_emiles} shows the residuals between SPAN and the tabulated E-MILES values for four commonly used diagnostic indices: H$\beta$, Mgb, Fe5270, and Fe5335.  
The agreement is excellent, with index-dependent residuals typically below $\sim$0.003\,\AA.  
These small offsets are attributable to discretisation effects and minor local inhomogeneities in the spectral resolution of the E-MILES templates.  
The achieved accuracy is approximately an order of magnitude better than the typical random error expected for line-strength indices at $\mathrm{S/N}\!\approx\!100$ ($\sim$0.02-0.05\,\AA; \citealt{cardiel1998}).

\begin{figure}
\centering
\includegraphics[width=14truecm]{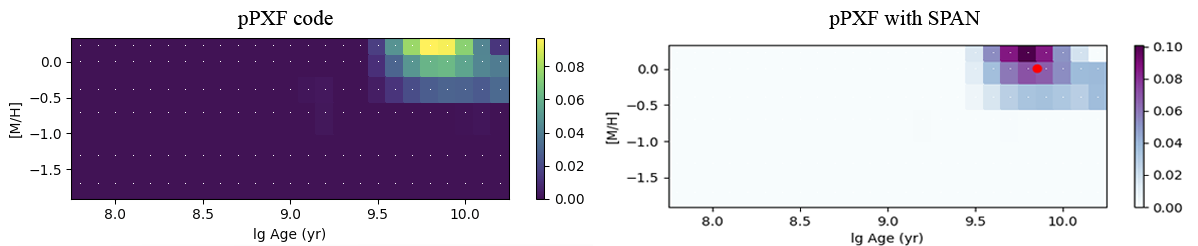}
\caption{Comparison of luminosity weights given by pPXF between the original example code provided by Cappellari (left panel) and SPAN (right panel) generated throughout the `Stellar population and SFH' task for the SDSS spectrum of NGC 3522.}
\label{fig:ppxf}
\end{figure}

For the `Stars and gas kinematics' and `Stellar populations and SFH' tasks, SPAN acts as a graphical wrapper that collects user inputs and passes the corresponding parameters to pPXF. Therefore, we first verified that the call to pPXF and the transmitted parameters were correctly handled and capable of reproducing well controlled tests. To this end, we compared the results of SPAN with the examples written by Cappellari and included in the pPXF distribution. The results given by SPAN are a perfect match both for the kinematics and stellar populations. \Cref{fig:ppxf} shows an example of the luminosity weights distribution for the SDSS DR8 spectrum of the galaxy NGC 3522 as fitted by the example code provided by Cappellari with the pPXF distribution (on the left) and with the `Stellar populations and SFH' task of SPAN (right plot), using the same parameters and the EMILES templates included with pPXF. The weights distribution, the mean age and metallicity values are exactly reproduced by SPAN.  These tests validate the GUI wrapper of SPAN.

As a futher step, we evauated SPAN in a full spectral analysis workflow by comparing its performances and results against the GIST pipeline \citep{bittner2019,bittner2021} using the publicly available example datacube\footnote{The GIST example working directory is publicly accessible at \url{https://abittner.gitlab.io/thegistpipeline/documentation/download/download.html}}. All tests have been performed on a Linux machine (Ubuntu 22.04) equipped with an Intel i5 processor and 4 GB of RAM. To ensure a reliable comparison, the same extraction and analysis parameters were used for both pipelines.

The extraction parameters were defined as follows: wavelength range of 4700--5600~\AA, redshift $z = 0.008764$, data origin at pixel coordinates (14, 14), no masking, a minimum S/N masking threshold of 5, and a target S/N of 100 for the Voronoi binning.  

For the GIST pipeline, we activated the following modules: stellar kinematics, gas emission lines (via GANDALF), and stellar populations. The spectral fitting was restricted to the 4800--5500~\AA \ region and the velscale value has been calculated by the \path{log_rebin} function of pPXF. We adopted a 8th-order additive polynomial for the kinematic fitting (with four Gauss--Hermite moments) and a 7th-order multiplicative polynomial for the stellar population fitting. The assumed noise level was 0.0163, and the regularisation was set using Regul. error = 0.04. Results have been computed in a luminosity-weighted framework. Parallelisation has been deactivated, and the full set of 625 EMILES SSP templates (BASTI isochrones, Kroupa IMF) was used.

To reproduce the equivalent analysis in SPAN, we first performed spectral extraction via the `Datacube Extraction' subprogram. Therefore, we loaded the generated spectra list followed by the activation and configuration of the `Stars and Gas Kinematics' and `Stellar Populations and SFH' tasks. For the `Stars and Gas Kinematics', we accounted for the MUSE Line Spread Function (LSF) by selecting the option `Spec. res. MUSE data' which automatically compute the MUSE LSF based on the Eq.8 of \citet{bacon2017}. We simultaneously fit the stellar and gas kinematics.
For the stellar population analysis, to match GIST configuration we computed the mean ages on a linear grid. 
The total execution time for the full workflow was comparable for both pipelines, taking approximately 1.5 hours.

\begin{figure}
\centering
\includegraphics[width=14truecm]{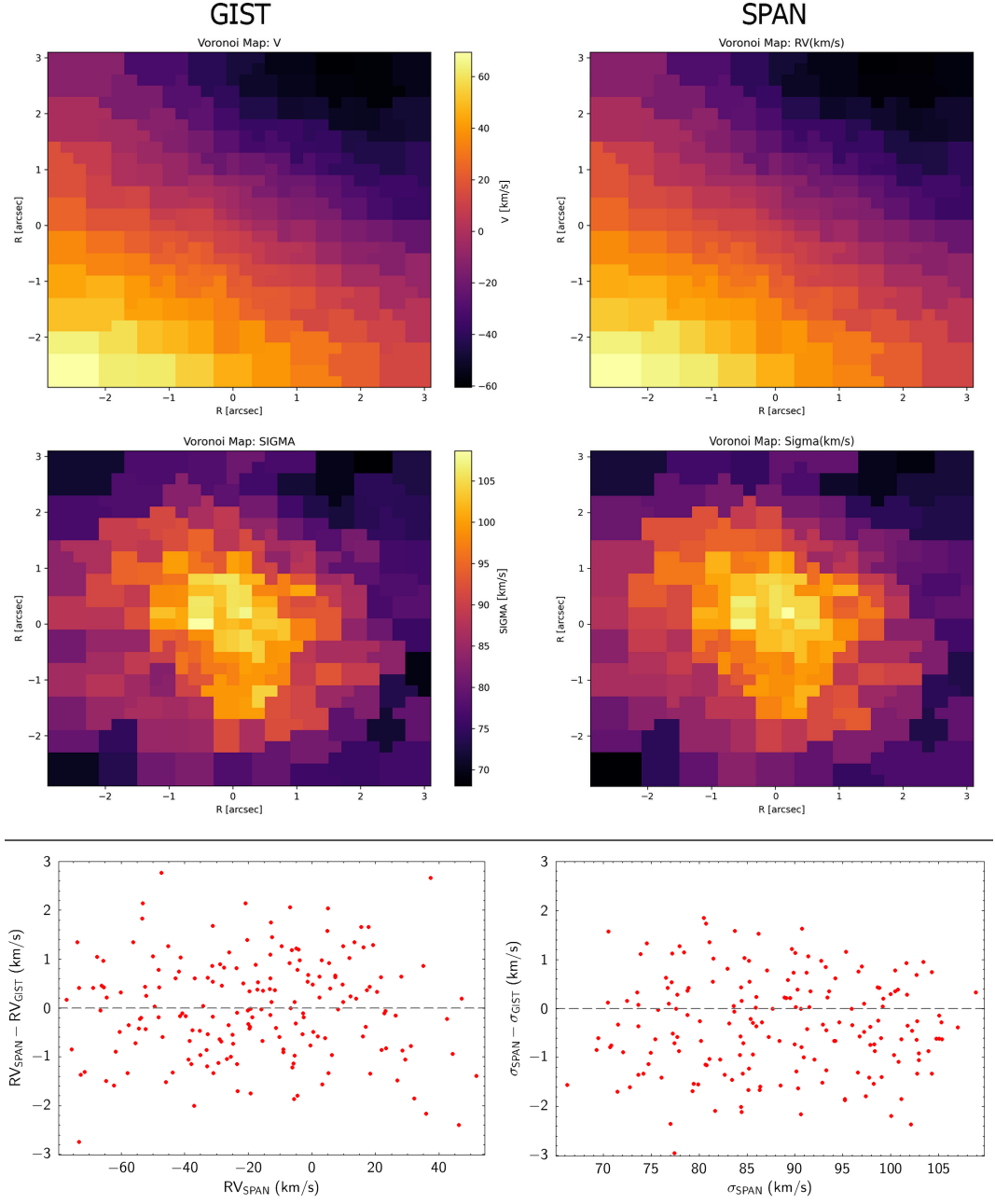}
\caption{Upper panels: Stellar kinematics maps computed with the `Plot maps' subprogram for the example datacube provided by the GIST pipeline. The left panels show the maps based on the analysis performed using the GIST pipeline and the right panels show the maps based on the analysis performed using SPAN. Lower panel: line-of-sight radial velocity (left) and velocity dispersion (right) residuals between SPAN and GIST as a function of the radial velocity and velocity dispersion calculated by SPAN. The dashed line marks the zero level value.}
\label{fig:gist-span}
\end{figure}

In \cref{fig:gist-span}, we present the comparison between the stellar kinematics derived by GIST and SPAN. The upper and middle panels show the line-of-sight velocity and velocity dispersion maps respectively, while the lower panels show the residuals between SPAN and GIST for the line-of-sight velocity and velocity dispersion. Both SPAN and GIST recovered the same pattern for the line-of-sight velocity and velocity dispersion, with no systematic offset detected for the line-of-sight velocity and a small $\sim$ 0.3 \kms \ offset for the line-of-sight velocity dispersion.

\begin{figure}
\centering
\includegraphics[width=14truecm]{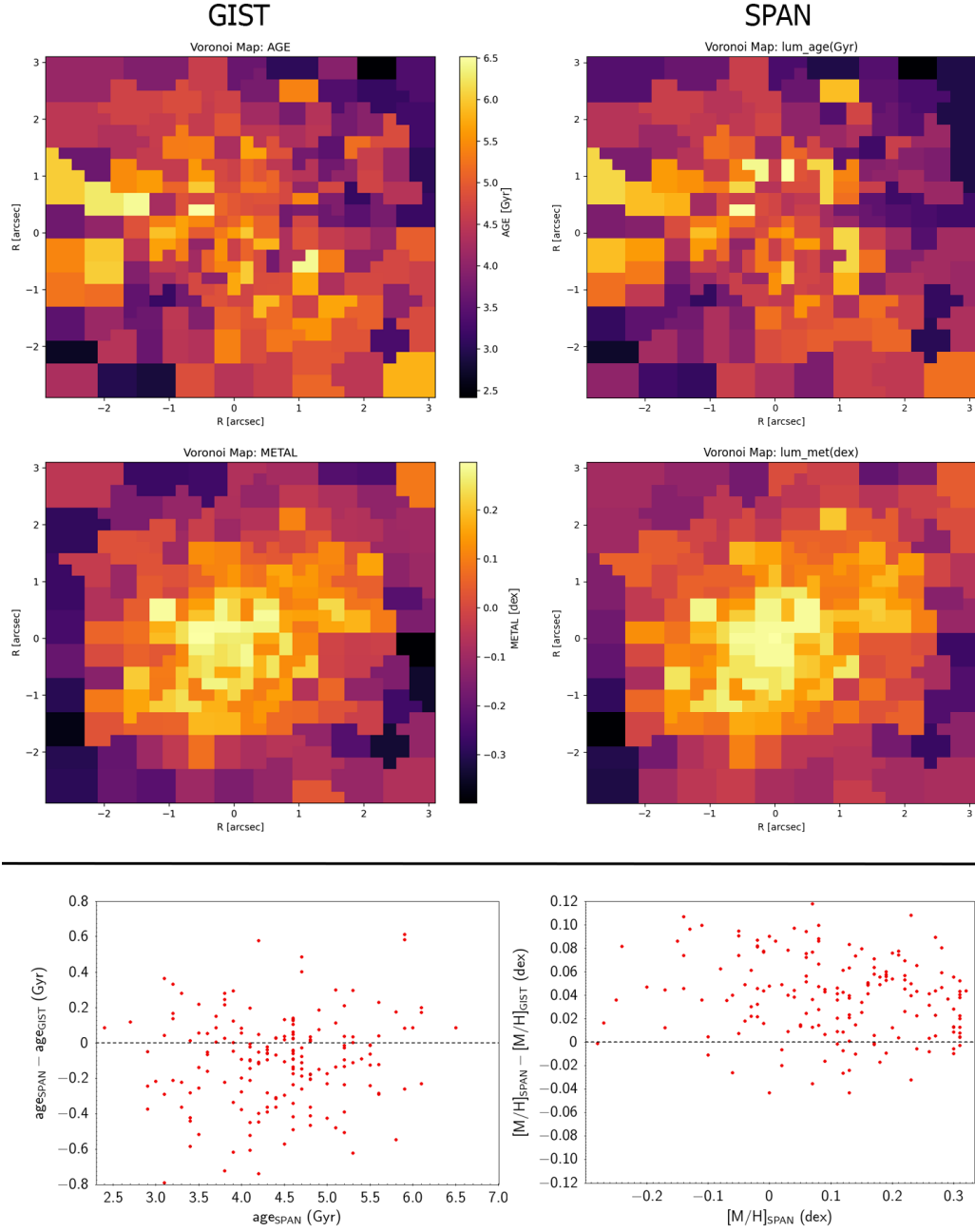}
\caption{Upper panels: Luminosity-weighted stellar parameter maps generated using the plotting subprogram of SPAN, applied to the example datacube provided by the GIST pipeline. The left panels show the maps based on the analysis performed using the GIST pipeline and the right panels show the maps based on the analysis performed using SPAN. Lower panel: age (left) and metallicity (right) residuals between SPAN and GIST as a function of the age and metallicity calculated by SPAN. The dashed line marks the zero level value.}
\label{fig:gist-span_sfh}
\end{figure}

\cref{fig:gist-span_sfh} shows the corresponding comparison for the luminosity-weighted stellar population parameters (age and metallicity). Despite methodological differences, particularly in the treatment of gas emission (performed by GANDALF for GIST and pPXF for SPAN), the results are in excellent agreement, with only minor deviations observed.

SPAN yielded on average slightly younger ages ($\sim$ 0.2 Gyr) and higher metallicities ($\sim$ 0.04 dex) with respect to GIST. This is likely due to the different emission line treatments between the two software. We did not observe this metallicity offset in the comparison with the example code of Cappellari using the SDSS spectra and the same parameters and templates (see \cref{fig:ppxf}).

\subsection{Uncertainties validation}
For all spectral analysis tasks, SPAN estimates the uncertainties associated with the derived quantities following standard approaches widely adopted in the literature such as Monte Carlo simulations for the EW measurements \citep[e.g.][]{morelli2008,morelli2015}, and for the kinematics with pPXF \citep[e.g.][]{morelli2008,mcdermid2015}, and bootstrap simulations for stellar population parameters with pPXF \citep[e.g.][]{kacharov2018,cappellari2023}. We tested the reliability of these uncertainty estimates. To this end, we relied on controlled simulations.

We selected an E-MILES BASTI template with solar metallicity and an intermediate age of 5 Gyr, representative of typical disc star-forming galaxies. Choosing an intermediate-age spectrum, rather than an old one ($\sim$10 Gyr), allows us to explore efficiently the main absorption features in the optical rest-frame range, including H$\beta$, which tends to be weaker in older populations.

From the reference template, we generated 100 simulated spectra by adding random Gaussian noise for six S/N levels, from 10 to 100. For each S/N value, we measured the EWs, stellar kinematics ($V$ and $\sigma$), stellar population parameters (luminosity-weighted age and metallicity), and their uncertainties as follows:

\begin{itemize}
    \item Equivalent widths (`Line-strength analysis' task):  
    The resolution of the reference template was first degraded from 2.51~\AA{} to 8.4~\AA{} to match the Lick/IDS system. For each S/N, we produced noisy realisations and measured the standard diagnostic indices H$\beta$, Mg$b$, Fe5270, and Fe5335. Uncertainties were obtained through 100 Monte Carlo simulations per spectrum and per index.
    \item Stellar kinematics (`Stars and gas kinematics' task):  
    The reference template was shifted by 200~km~s$^{-1}$ and broadened by 150~km~s$^{-1}$ to simulate realistic kinematic parameters before noise addition. We fitted the 4800--5500~\AA{} spectral region using the E-MILES SSP subset provided by pPXF, with an additive polynomial of degree~4. For each simulated spectrum, 100 Monte Carlo iterations were performed to estimate the uncertainties.
    \item Stellar populations (`Stellar populations and SFH task'):  
    Using rest-frame, unbroadened noisy simulated spectra, we fitted the same 4800--5500~\AA{} wavelength range, adopting standard pPXF parameters for medium-S/N ($\sim$50) spectra with good flux calibration \citep[see][]{kacharov2018}: regularisation $=50$ (regul. error $=0.02$), and low-order multiplicative polynomial of degree~3. For each simulated spectrum, 100 bootstrap realisations were used to estimate the uncertainties.
\end{itemize}

\begin{figure}
\centering
\includegraphics[width=14truecm]{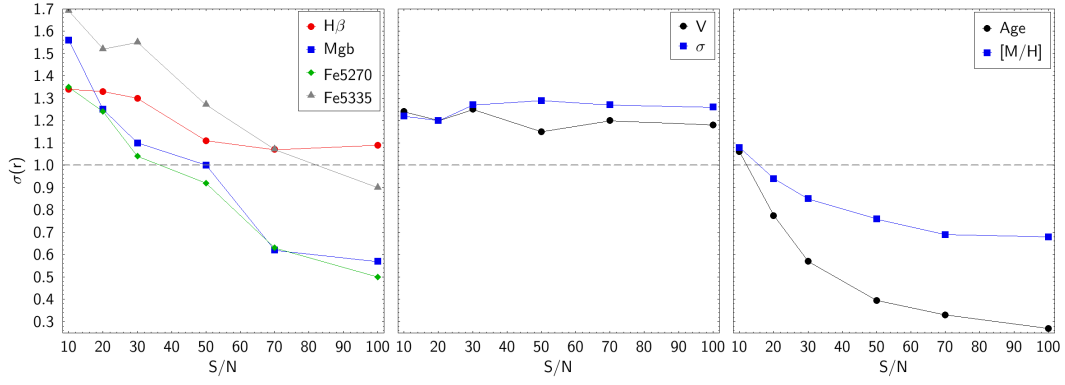}
\caption{Error normalised standard deviation as a function of the S/N for the EW, kinematics and stellar populations calculated from 100 simulated noisy spectra for each S/N value.}
\label{fig:error_analysis}
\end{figure}

For each realisation and S/N level, we computed the error-normalised residuals:
\[
r = \frac{Q_{\mathrm{i}} - Q_{\mathrm{true}}}{\sigma_{Q}}
\]
where $\sigma_{Q}$ is the $1\sigma$ uncertainty estimated by SPAN for the measured quantity $Q_{\mathrm{i}}$, compared with the true value $Q_{\mathrm{true}}$.  
For each S/N bin, we then calculated the standard deviation of the error-normalised residuals, $\sigma(r)$, and plotted this quantity as a function of S/N. The results are presented in \cref{fig:error_analysis} for the EW, kinematics and stellar populations respectively. We find that:

\begin{itemize}
    \item Equivalent widths (Monte Carlo):  
The error-normalised dispersions $\sigma(r)$ show index-dependent trends with S/N. This behaviour is a direct consequence of how the EW uncertainties are determined. In SPAN, the noise amplitude for the Monte Carlo realisations is inferred empirically from the rms of the residuals in the pseudo-continuum bands of each index, following the common practice adopted in line-strength studies when only local S/N information is available \citep[e.g.][]{cardiel1998,cenarro2001,cesetti2009}. At S/N~$\gtrsim 50$, the rms measured in the pseudo-continuum bands is no longer dominated by random noise, but also by intrinsic small-scale spectral structure (line blanketing and curvature with respect to a straight-line continuum). This introduces an effective noise floor that does not scale with S/N and naturally leads to mildly conservative, index-dependent uncertainties.  
At low S/N (~$\lesssim 20$), the pseudo-continuum bands are dominated by noise, but the current Monte Carlo implementation perturbs only a synthetic continuum within the index definition bands and does not propagate the uncertainty in the continuum placement itself. As a consequence, in this low S/N regime the uncertainties returned by SPAN can be underestimated by up to a factor of $\sim$1.5-1.7 for the weakest indices. We note, however, that S/N~$\lesssim 20$ lies at or below the thresholds commonly adopted for reliable index measurements, and that additional systematic effects are known to arise in this regime \citep[e.g.][]{morelli2020,morelli2025}.

    \item Stellar kinematics (Monte Carlo):
    Velocity and dispersion residuals show stable dispersions, with $\sigma(r_V)\approx1.15{-}1.25$ and $\sigma(r_\sigma)\approx1.20{-}1.30$ over S/N~=10--100, and no significant S/N dependence. This indicates statistically well-calibrated uncertainties, with only a mild ($\lesssim$20\%) deviation from unity.
    \item Stellar populations (Bootstrap): 
    For luminosity-weighted parameters, $\sigma(r_{\mathrm{age}})$ is mildly conservative at high S/N ($\sim$0.3--0.4 at S/N $\geq$ 50) and approaches unity as S/N decreases ($\sim$1.1 at S/N $=10$).  
    For metallicity, $\sigma(r_{\mathrm{[M/H]}})$ is slightly conservative at high S/N ($\sim$0.7) and close to unity at lower S/N ($\sim$1.0--1.1 for S/N $\sim$ 10--20).      
    This trend may be a consequence of the residual-bootstrap approach applied to a non-linear, degenerate problem: at low S/N the fit residuals are dominated by the Gaussian noise and the bootstrap closely tracks the true noise-driven scatter ($\sigma(r)\approx1$), whereas at high S/N the residuals are increasingly shaped by template limitations, age-metallicity degeneracy, regularisation, and continuum-shape polynomials, which do not scale with 1/(S/N) and are interpreted by the bootstrap as additional variance. 

\end{itemize}

\section{Example science case: the spiral galaxy NGC 1097}
\label{sec:examples}

In this section, we show how to use SPAN to efficiently analyse spectroscopic data and produce science-grade results. As a representative case study, we focus on the kinematics and stellar populations of the spiral galaxy NGC~1097. This galaxy has been observed as part of the Time Inference with MUSE in Extragalactic Rings (TIMER) project \citep{gadotti2019}, with analyses presented in \citet{gadotti2019} and \citet{bittner2020}.

The fully reduced MUSE datacube of NGC~1097 was retrieved from the ESO Science Archive. Using the `Datacube extraction' subprogram, we followed the methodology described in \citet{bittner2020}: we generated Voronoi bins with a target S/N of 100 per pixel, and masked all spaxels with S/N $\leq$ 3. To optimise execution time, we restricted the spectral extraction to the 4750 -- 5800~\AA\ range, and we estimated the S/N on the same wavelength range. Finally, we set the redshift to z = 0.004240\footnote{The redshift value has been retrieved from the NASA/IPAC Extragalactic Database: \url{https://ned.ipac.caltech.edu/}} to extract de-redshifted spectral bins. The `Datacube extraction' produced 4175 spectral bins, automatically listed in the spectra list file, which was then loaded into the main SPAN interface.

\subsection{Stars and gas kinematics}
\label{sec:optical_analysis}

\begin{figure}
\centering
\includegraphics[width=10truecm]{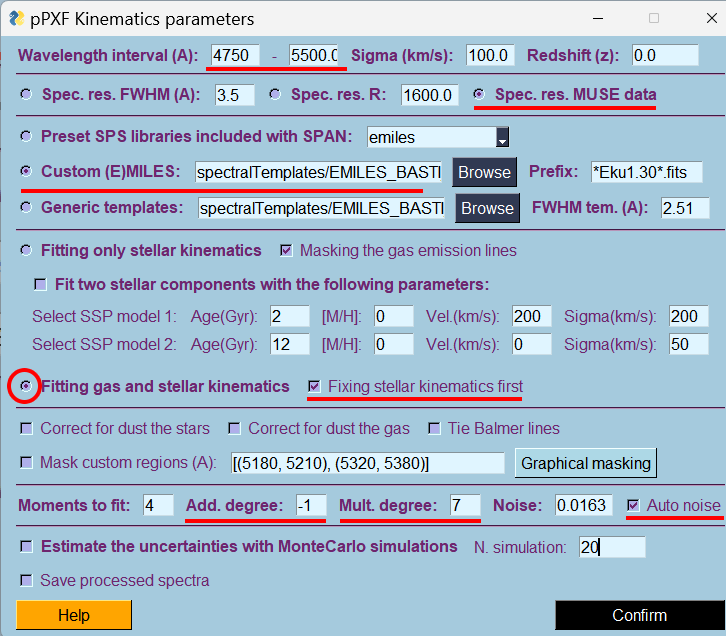}
\caption{Parameter panel for the `Stars and gas kinematics' task in SPAN. Parameters modified from their default settings are marked in red to reproduce the workflow described by \citet{gadotti2019}.}
\label{fig:kin_params}
\end{figure}

For the kinematic analysis, we activated the `Stars and gas kinematics' task and fine-tuned the parameters to pass to pPXF following the indications of \citet{gadotti2019}. We employed the full set of E-MILES BASTI templates with a Kroupa IMF, available in the \path{SpectralTemplates} folder of SPAN, and limited the fitting range to 4750 -- 5500~\AA.
Since the spectral resolution of MUSE data is a function of the wavelength, we enabled the `Spec. res. MUSE data' option, which automatically computes the MUSE LSF. To account for small differences in the continuum shape of the extracted spectra, we set the degree of the multiplicative Legendre polynomial to 7 and disabled the additive polynomial (set to -1). We fitted four moments of the line-of-sight velocity distribution (LOSVD),  and selected the `Fitting gas and stellar kinematics' with the option 'Fixing stellar kinematics first'. This allowed us to extract both the stellar kinematics (with emission lines masked) and the gas kinematics by fixing the LOSVD moments of the stellar component. Finally, we activated the `Auto noise' option to let pPXF calculate the mean noise of the spectra and give a realistic representation of the uncertainties without relying on computationally expensive Monte Carlo simulations.
\Cref{fig:kin_params} shows the parameter panel for the `Stars and gas kinematics' task. Parameters modified from their default values for this analysis are highlighted in red. It is worth noting that these are the parameters required and handled by pPXF. Different choices of fitting parameters or templates are expected to have only a minor impact on the derived stellar kinematics, as also reported by \citet{gadotti2019,gadotti2020,cappellari2023}.

\begin{figure}
\centering
\includegraphics[width=14truecm]{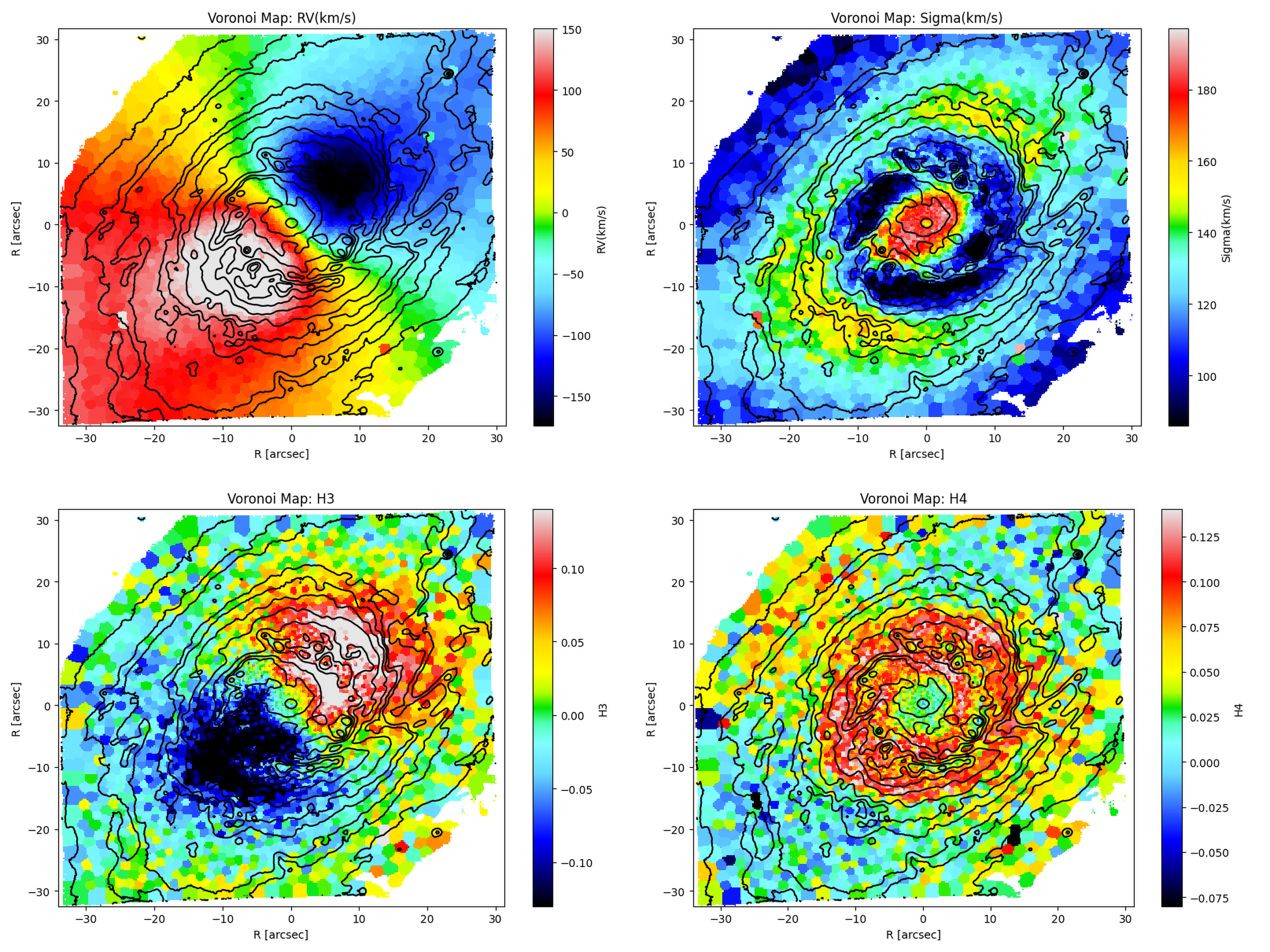}
\caption{Stellar kinematic maps of NGC~1097 generated with SPAN, showing the line-of-sight velocity (V$_{*}$), velocity dispersion ($\sigma_{*}$), h$_3$, and h$_4$ moments.}
\label{fig1097_muse_kin_stars}
\end{figure}

The resulting maps of line-of-sight stellar velocity, velocity dispersion, and higher order moments (h$_3$ and h$_4$) are shown in \cref{fig1097_muse_kin_stars}. These have been generated using the `Plot maps' subprogram. 
We notice strong visual similarities with the maps presented in \citet[][see their Fig.~3]{gadotti2019}. 
SPAN successfully reproduces all the main dynamical features identified by the TIMER team (see their Sec.~4.2), which can be summarised as follows:
 
\begin{itemize}
\item A rapidly rotating inner disc with V$_{*} \sim 150$~km~s$^{-1}$;
\item A two-zone structure in velocity dispersion ($\sigma_{*}$): an inner region ($\sim$10 arcsec) with low $\sigma_{*}$, and a surrounding region ($\sim$15 arcsec) with significantly higher values;
\item Peak $\sigma_{*}$ values distributed in the central region ($\sigma_{*} \sim 180$~\kms);
\item A strong anti-correlation between V$_{*}$ and h$_3$, indicative of near-circular orbits in the inner disc;
\item High h$_4$ values for the inner disc, which suggest the superposition of structures with different LOSVD.
\end{itemize}

\begin{figure}
\centering
\includegraphics[width=14truecm]{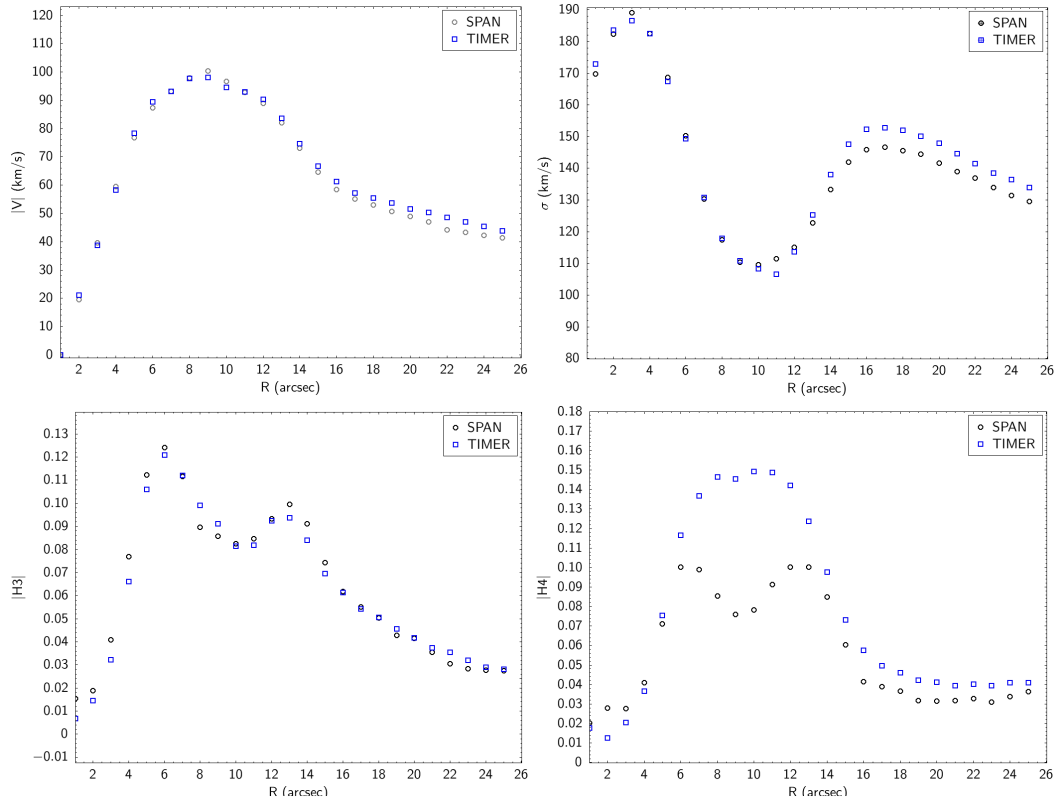}
\caption{Radial profiles of line-of-sight velocity, velocity dispersion, h$_3$, and h$_4$ moments between SPAN (black circles) and the TIMER results from \citet{gadotti2020} (blue squares) for 1 arcsec wide elliptical annuli.}
\label{fig:profiles_kin}
\end{figure}

To perform a quantitative validation and to minimise the impact of the different Voronoi samplings between the SPAN and TIMER datasets, we downloaded the stellar kinematics maps computed by \citet{gadotti2020} from the TIMER website\footnote{\url{https://www.muse-timer.org/data}} and we extracted one-dimensional radial profiles by averaging the kinematic quantities within concentric elliptical annuli, both for SPAN and TIMER maps. We defined 25 annuli, each 1\,arcsec wide, centred on the photometric nucleus of the galaxy, with ellipticity q\,=\,0.84 and position angle (PA\,=\,144$^{\circ}$) following the mean isophotal profile. The resulting profiles are shown in \cref{fig:profiles_kin}.  
For the line-of-sight velocity, h$_3$, and h$_4$, we used absolute values when constructing the radial profiles to avoid spurious fluctuations in the mean trends caused by opposite signs and by the different Voronoi samplings between SPAN and TIMER.

The distributions obtained from SPAN and TIMER are in excellent agreement, with only minor local deviations that do not affect the global trends. The median residuals (SPAN--TIMER) are $-1.72$\,km\,s$^{-1}$ for the velocity, $-3.17$\,km\,s$^{-1}$ for the velocity dispersion, $+0.0007$ for $h_3$, and $-0.01$ for $h_4$.  
Overall, the SPAN kinematic measurements closely reproduce the TIMER results, with small differences, particularly for $h_4$, which likely arise from different Voronoi sampling ($\sim$ 30000 bins for TIMER and $\sim$ 4000 for SPAN), adopted template libraries, and pPXF configuration parameters  (e.g. the regularisation factor, which is unknown for TIMER analysis).

\subsection{Stellar populations and SFH}
\label{sec:stellar_pop}

For the stellar population analysis, we used the same Voronoi spectral bins and activated the `Stellar populations and SFH' task with the `Fitting stars and gas together' option, allowing pPXF to simultaneously model the emission lines and fit the emission-corrected spectra. We adopted the configuration of \citet{bittner2020} as our reference setup, with minor adjustments. After several tests, we restricted the original 4800--5800 \AA\ wavelength range used by \citet{bittner2020} to 4800--5580 \AA, because pPXF struggled to properly model the emission lines over the wider range, leading to artificially older ages (the authors used GANDALF and a two step method to model the emission lines).

\begin{figure}
\centering
\includegraphics[width=10truecm]{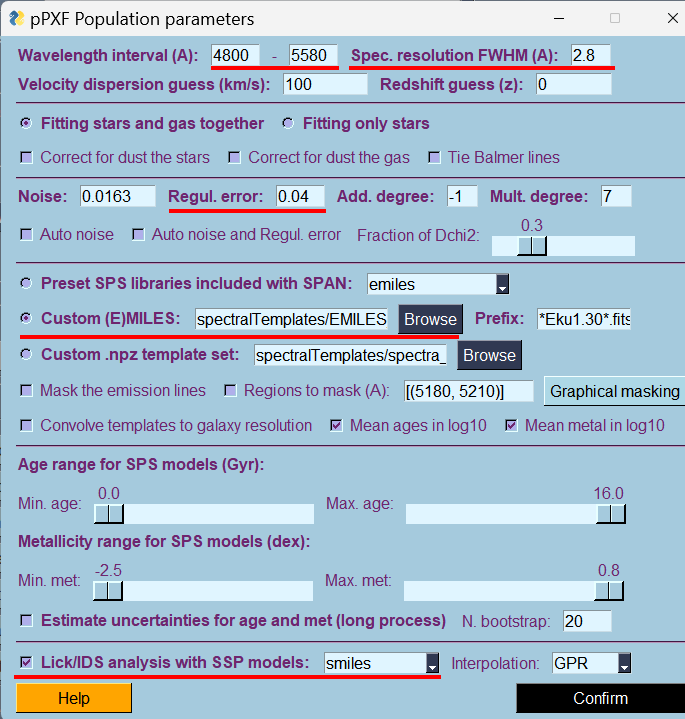}
\caption{Parameter panel for the `Stellar populations and SFH' task in SPAN. Parameters modified from their default settings are marked in red to reproduce the workflow described by \citet{bittner2020}.}
\label{fig:sfh_params}
\end{figure}

To speed up the process, we only considered age and metallicity values, maintaining the same template set of EMILES SSPs with BASTI isochrones and base $[\alpha/Fe]$ of the kinematics analysis. We set the resolution to a mean value in the considered wavelength range (FWHM = 2.8 \AA), and the multiplicative Legendre polynomial of pPXF to degree 7. We fixed the mean noise of each spectrum to 0.0163, and we adopted a fixed regularisation (`Regul. error' = 0.04) for each bin. As noted by \citet{bittner2020}, the choice of regularisation and noise has little impact on the mean age and metallicity values, while \citet{cappellari2023} showed that the solution becomes stable for non-zero polynomial degrees, making the exact polynomial degree selection non-critical.
\Cref{fig:sfh_params} shows the parameter panel for the `Stellar populations and SFH' task. Parameters modified from their default values for this analysis are highlighted in red.

\begin{figure}
\centering
\includegraphics[width=14truecm]{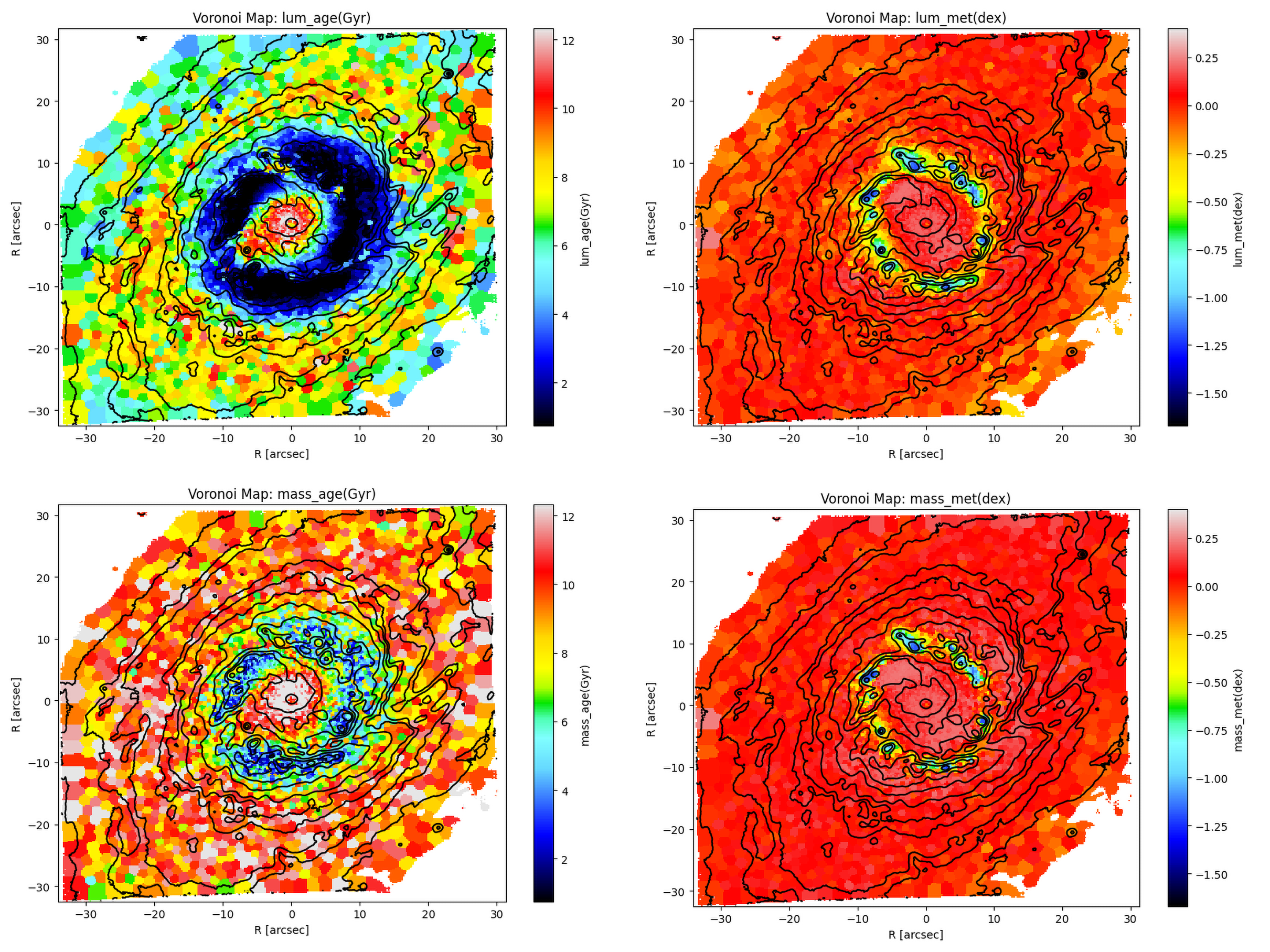}
\caption{Luminosity-weighted mean stellar age and metallicity maps of NGC~1097, derived from full spectral fitting with SPAN.}
\label{fig:muse_pop}
\end{figure}

\Cref{fig:muse_pop} shows the luminosity (upper panel) and mass (lower panel) weighted mean stellar age and metallicity maps obtained through full spectral fitting with pPXF, generated using the `Plot maps' subprogram. Despite the differences in template sets and the exclusion of $[\alpha/\mathrm{Fe}]$ parametrisation, our results visually show all the small-scale features detected by  \citet[][see their Fig.~4, upper panels]{bittner2020}. In particular, the nuclear disc appears as the oldest and most metal-rich region in both luminosity and mass weighted maps. In contrast, the surrounding star forming nuclear ring is characterised by young and metal poor stellar populations, even when considering mass-weighted values.

\begin{figure}
\centering
\includegraphics[width=14truecm]{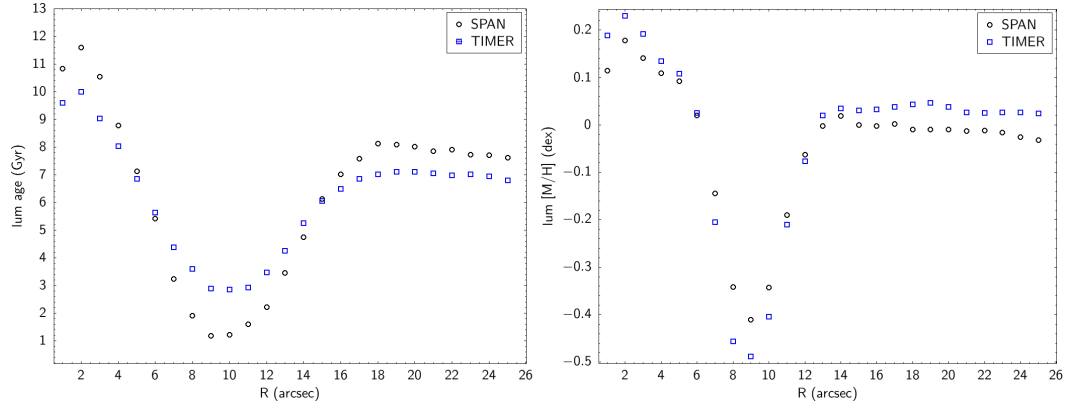}
\caption{Radial profiles of luminosity weighted age (left) and metallicity (right) between SPAN (black circles) and the TIMER results presented in \citet{bittner2020} (blue squares) for 1 arcsec wide elliptical annuli.}
\label{fig:profiles_sfh}
\end{figure}

To perform a quantitative validation, we applied the same approach of \cref{sec:optical_analysis} to derive elliptical radial profiles of the luminosity-weighted age and metallicity from both SPAN and TIMER maps. The resulting profiles are shown in \cref{fig:profiles_sfh}.  
The two datasets exhibit very good overall agreement, reproducing the same radial gradients and global trends.  Nonetheless, some differences are observed, particularly for the luminosity-weighted age: SPAN results span a wider age range, yielding slightly older values at the high end and younger ones at the low end. The median residual (SPAN--TIMER) is $-0.71$\,Gyr, indicating that SPAN provided marginally younger ages on average.  
For metallicity, the agreement is even closer, with SPAN returning slightly lower values in the solar to supersolar regime and slightly higher values in the sub-solar range, with a median residual of $-0.03$\,dex.

The small differences observed between SPAN and TIMER are consistent with the level of systematic uncertainties inherent in full spectral fitting analyses. They primarily reflect well-known sensitivities to the adopted stellar population models, assumptions on [$\alpha$/Fe], regularisation level, continuum polynomials, and wavelength range, rather than any SPAN-specific bias \citep[see also][]{bittner2020}.  
More specifically:
\begin{itemize}
    \item The use of stellar population models with or without an explicit [$\alpha$/Fe] parametrisation \citep[see Section~4.1 of][]{bittner2020};
    \item The modelling accuracy of the gaseous emission component (Section~4.2 of the same work);
    \item In our specific case, the direct treatment of emission lines within pPXF, as opposed to the two-step procedure adopted by the TIMER team, which first employed GANDALF to model and subtract emission lines before fitting the stellar component.
\end{itemize}

\begin{figure}
\centering
\includegraphics[width=14truecm]{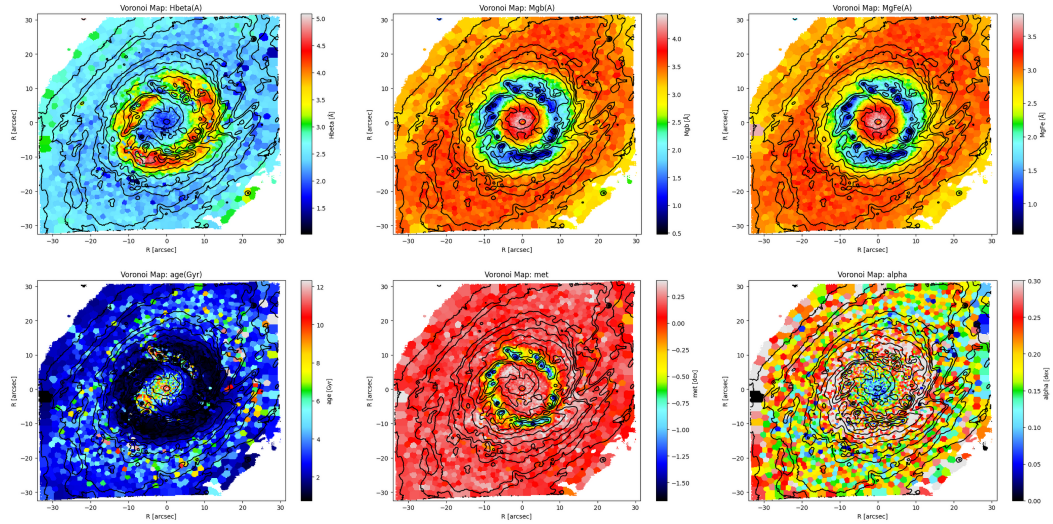}
\caption{Lick/IDS H$\beta$, Mgb, and [MgFe]$^{`}$ indices (upper panel) and luminosity weighted age, metallicity and $[\alpha/Fe]$ (lower panel) estimated using the sMILES SSP models and the GPR algorithm.}
\label{fig:muse_lick}
\end{figure}

For further comparison, in \cref{fig:muse_lick} we show the results of the Lick/IDS analysis. The upper panels display the EWs of the stellar H$\beta$, Mgb, and [MgFe]$^{\prime}$ indices, while the lower panels show the corresponding mean stellar ages, metallicities and $[\alpha/\mathrm{Fe}]$.

Compared to the full spectral fitting, the Lick/IDS-based estimates tend to yield systematically younger ages and higher metallicities. This discrepancy likely arises from the simplified assumption of SSP inherent in the index-based method. Nevertheless, the Lick/IDS analysis reveals spatial trends that are consistent with those derived from pPXF, providing an independent and complementary validation of the robustness of the results.

\section{Future Developments}
\label{sec:future}
SPAN is an actively maintained and evolving project. Several key developments are planned to further enhance its scientific capabilities and broaden its applicability across a wider range of spectroscopic datasets and research contexts.

\begin{itemize}
\item Expansion of the datacube direct support:
The current version of SPAN provides dedicated extraction routines for datacubes from the MUSE, CALIFA, WEAVE LIFU, and JWST NIRSpec IFU instruments. While the software is already capable of handling datacubes from any instrument through user-defined routines, future releases will expand the library of built-in modules to include direct support for additional instruments. This development aims to minimise the need for custom coding, speed up the extraction process, and provide a more complete user experience when working with datacubes.
\item Integration of additional full spectral fitting algorithms:  
While pPXF currently serves as the core engine for full spectral fitting, future versions of SPAN will incorporate additional spectral fitting algorithms to provide users with a broader range of modelling approaches. This will allow direct comparison between different methods and will offer greater flexibility. 
\item Fine-tuning and further testing of the GPR method for line-strength analysis:  
The application of GPR to stellar population parameter estimation via Lick/IDS index analysis and SSP model comparison has yielded promising results, offering a significant speed-up compared to standard multi-dimensional interpolation techniques. We will conduct more extensive testing to better evaluate the accuracy and robustness of the method, and we will optimise the training procedure and parameter settings for each adopted SSP model grid.

\end{itemize}

\section{Conclusion}
\label{sec:conclusion}
In this paper we presented SPAN, a cross-platform GUI software written in Python (compatible with Python 3.10 and above), designed for the manipulation and analysis of astronomical spectra, optimised for the optical to NIR wavelength range and for galaxy spectra.

SPAN has been developed as a long-term project with the aim of providing the community with a cross-platform, intuitive and accessible GUI that consolidates the most widely used tools for modern spectral analysis. Its primary goal is to simplify the workflow of spectral processing and interpretation by integrating advanced techniques, such as line-strength and full spectral fitting with pPXF, into a single, coherent environment. This significantly reduces the learning curve required to access and manage state-of-the-art methodologies used in spectroscopy.

The software includes both built-in routines and wrappers to widely adopted external libraries and algorithms such pPXF. The GUI has been optimised for compatibility across major operating systems, including Linux, Windows, macOS, and Android. The interface has been developed using the FreeSimpleGUI wrapper for the Python \path{Tkinter} framework.

Validation tests show that the outputs produced by SPAN are fully consistent with those obtained using well established codes and pipelines typically employed in spectral analysis workflows.

The SPAN distribution includes sample spectra and a complete user manual, providing detailed documentation of all available tasks and configurable parameters.

\section{Acknowledgements} 
We thank the anonymous referee for the constructive and insightful comments that helped improve the quality of this paper.\\
We thank Michele Cappellari for providing useful suggestions, and the permission to use and adapt part of his code to SPAN. \\
DG acknowledges the support from Comité Mixto ESO-Gobierno de Chile Grant no. ORP060/19.\\
DG and LM acknowledge the support from PROYECTOS  FONDO  de ASTRONOMIA ANID -  ALMA  2021 Code :ASTRO21-0007. \\
UB is supported through the Italian Ministry of Research (FARE project EASy R20SLAA8CJ).\\
JMA acknowledges the support of the Agencia Estatal de Investigación del Ministerio de Ciencia e Innovación (MCIN/AEI/10.13039/501100011033) under grant nos. PID2021-128131NB-I00 and CNS2022-135482 and the European Regional Development Fund (ERDF) ‘A way of making Europe’ and the ‘NextGenerationEU/PRTR’\\
AdLC  acknowledges financial support from the Spanish Ministry of Science and Innovation (MICINN) through RYC2022-035838-I and PID2021-128131NB-I00 (CoBEARD project).

%%%%%%%%%%%%%%%%%%%%%%%%%%%%%%%%%%%%%%%%%%%%%%%%%%%
\section{Data Availability} 
SPAN is publicly available on PyPi (\url{https://pypi.org/project/span-gui/}) and on GitHub: (\url{https://github.com/danielegasparri/span-gui}) and can be installed via pip. Additional documentation including video tutorials are available on  the personal web page of the author (\url{https://www.danielegasparri.com/span-software/}).

The example MUSE datacube of NGC 1097 is publicly available on ESO Science Portal. 

\newpage

%\appendix

\bibliographystyle{elsarticle-harv.bst}
\clearpage
\bibliography{bib_span}

\end{document}